\documentstyle[aps,prd,tighten]{revtex}

\begin{document}
\draft

\title{$< T^{\mu}_{\nu}>_{ren}$ of the quantized conformal fields 
 in the Schwarzschild spacetime: Israel - Hartle - Hawking state}
\author{Jerzy Matyjasek}
\address{Institute of Physics, Maria Curie Sk\l odowska University\\
pl. Marii Curie Sk\l odowskiej 1, 20-031 Lublin, Poland\\
email: matyjase@tytan.umcs.lublin.pl\\
jurek@iris.umcs.lublin.pl }
\maketitle

\begin{abstract}
The renormalized expectation value of the stress energy  tensor
of the conformally invariant massless fields
in the Israel-Hartle-Hawking  state in the Schwarzschild spacetime is constructed.
It is achieved through solving the conservation equation
in conformal space and utilizing the regularity  conditions in a physical
metric. Specifically, the relation of the results of the present approach to the
stress tensor constructed within the framework of the Hadamard renormalization 
is analysed. Finally, the semi-analytic models reconstructing the 
numerical estimates of the tangential component of the stress-energy tensor with
the maximal deviation not exceeding $0.7 \%$ are constructed.

\end{abstract}

\bigskip \noindent{04.62+v, 04.70.Dy \\UMCS-FM-98-17\\August 1998}
\bigskip
\par
\baselineskip=18pt
In a recent publication~\cite{J1} we have constructed the approximate mean 
value of the regularized
stress-energy tensor of the massless and conformally invariant quantized scalar field 
in the Israel-Hartle-Hawking state in the Schwarzschild spacetime.
We  employed the Hadamard regularization which has proven to be 
a powerful tool in such calculations~\cite{Brown1}-~\cite{Mario}. 
In the Hadamard regularization one must solve the constraint equations for three
unknown functions of  radial coordinate
and insert them into the general expression for $<T^{\mu}_{\nu}>~\cite{Tadaki1}.$
Unfortunately, since  the constraint equation involves three unknown functions,
one of which being closely related to the vacuum fluctuation of the quantized field
$<\phi^{2}>,$  the problem, beside the boundary conditions, must be supplemented by
some additional informations regarding their nature~\cite{Steve}, 
\cite{Candelas1}.

In Ref. [1] we have assumed that unknown functions (and $<\phi^{2}>$) have 
a simple form 
\begin{equation}
\sum^{k^{\prime}}_{m=-k} ( 1\,-\,x)^{m}\,W_{m}(x), \hspace{1in}
k,\,k^{\prime}\,\geq 0,
\end{equation}
where $x\,=\,2M/r,$ and $W_{m}(x)$ for each value of $m,$ is a polynomial in 
$x,$ and showed that it is relatively easy to construct solutions, which lead
to the stress-energy tensor and the vacuum fluctuation which reflects
principal features of the exact $<\phi^2>_{ren},$ and $<
T^{\mu}_{\nu}>_{ren}$ with a great accuracy. 
Resulting three-parameter stress-energy tensor may be further determined
from the known horizon value of the one of the components of  $<
T^{\mu}_{\nu}>_{ren}$ and making use of the equation
\begin{equation}
<T^{r}_{r}(2 M)>_{ren}\,=\,{\kappa\over 6}{d \over dr}<\phi^2 >_{ren}\,+\,{16\over 15}\pi^2
T_{H}^{4},
\end{equation}
where $\kappa$ is the surface gravity and $T_{H}$ stands for the black hole temperature.
The last unknown parameter has been fixed by some sort of a best fit argument.

In this note we shall show how the results of Ref. [1] may be obtained 
and generalized in a more systematic
and simpler way, without recourse to $<\phi^2>_{ren}.$ 
Moreover,  a great advantage of the  adopted method is that it
could be, contrary to the Hadamard regularization, easily extended 
to conformally invariant massless spinor and vector fields. 
Our present approach is based on the method adopted earlier in the different context
and uses scaling properties of the one-loop renormalized effective action 
under the conformal transformations, or more precisely their consequences
for appropriate transformations of the renormalized 
stress-energy tensors~\cite{Brown2}-\cite{Zannias}. The notation
is essentially that of Refs.~\cite{J2}\cite{J5} , to which the reader 
is referred for details. 
In this method, employing the Christensen-Fulling
asymptotic analyses~\cite{Steve}, one assumes that the 
tangential component of the stress-energy tensor in the optical companion to the 
Schwarzschild space has a simple polynomial form
\begin{equation}
<\tilde{T}_{\theta }^{\theta }>_{ren}=T p(s)\sum_{n=0}^{N}a_{n}x^{n},
\end{equation}
with $a_{0}\,=\,1,$ 
where 
$T=\pi^{2}T_{H}^{4}/90$ and
$p(s)$ is a numerical coefficient depending on the spin of the field.
Here $p(0) = 1,$ $p(1/2) = 7/4,$ and $p(1) = 2.$
We distinguish quantities
evaluated in the conformal space by a tilde.
Subsequently, solving the conservation equation in the conformal space for the radial component
of the stress tensor and utilizing regularity conditions on the event horizon 
in the physical space one 
reduces the number of unknown coefficients $a_{i}.$ 
Their number may be further substantially
reduced accepting one of the two thermal hypotheses,
which state that the stress tensor in the Israel-Hartle-Hawking vacuum
should have the form
\begin{equation}
< T_{\nu }^{\mu }>_{ren}\,=\,p(s) T \left[1\,+\,\sum_{n=1}^{m}(n\,+\,1) \,x^{m}\right]
(\delta^{\mu}_{\nu}\,-\,4 \delta^{\mu}_{0}\delta^{0}_{\nu})
\,+\,O(x^{m+1}),
\end{equation}
with $m\,=\,2$ for the weak thermal hypothesis and $m\,=\,5$ for the strong one~\cite{Visser}.
The weak thermal hypothesis is usually motivated by the observation that since the curvature 
is proportional to $x^{3}$ the curvature corrections to the stress-energy tensor
are expected to be of that order. On the other hand, in the strong version
of this ansatz one assumes that the curvature corrections are proportional to $x^{6},$
i. e. $(\it{curvature}^{2}).$
In practice, it is helpful to invert the order of operations, and to analyse the consequences
of the regularity conditions and thermal hypotheses imposed on the tangential 
component in the Schwarzschild geometry
before  solving the conservation equation.

The stress-energy tensor under the conformal transformation, 
${\tilde g}^{\mu\nu}\,=\,\exp (-2\omega) g_{\mu\nu},$
transforms as
\begin{equation}
< T_{\nu }^{\mu }>_{ren} \,=\,\exp {(-4\omega )}\,{\tilde{T}}_{\nu
}^{\mu }\,+a(s)A_{\nu }^{\mu }\,+\,b(s)B_{\nu }^{\mu }\,+\,c(s) C_{\nu}^{\mu},
\end{equation}
where 
\begin{equation}
A^{\mu \nu }\,=\,8R^{\alpha \mu \nu \beta }\omega _{;\alpha \beta }-{\frac{4%
}{3}}\kappa ^{;\mu \nu }+2g^{\mu \nu }\left( 2\omega ^{;\alpha }\kappa
_{;\alpha }\,+\,\kappa ^{2}\,+\,{\frac{2}{3}}\Box \kappa \right)
\,-\,8\kappa ^{;(\mu }\omega ^{;\nu )}\,-\,8\omega ^{;\mu }\omega ^{;\nu
}\kappa ,
\end{equation}
\begin{eqnarray}
B^{\mu \nu }\, =\,8R^{\alpha \mu \nu \beta }\omega _{;\alpha \beta
}\,+\,8R^{\alpha \mu \nu \beta }\omega _{;\alpha }\omega _{;\beta
}\,-8\omega ^{;\mu \alpha }{\omega _{;\alpha }}^{;\nu }\,-\,8\kappa ^{;(\mu
}\omega ^{;\nu )}\,-\,8\kappa \omega ^{;\mu }\omega ^{;\nu }\,  \nonumber \\
\quad +\,4g^{\mu \nu }\left( \omega _{;\alpha \beta }\omega ^{;\alpha
\beta }\,+\,\kappa _{;\alpha }\omega ^{;\alpha }\,+\,{\frac{1}{2}}\kappa
^{2}\right) ,
\end{eqnarray}
\begin{eqnarray}
C^{\mu\nu}\,=\, g^{\mu\nu}(2\Box\kappa\,+\,3\kappa^2\,+\,6 \omega_{;\alpha}
 \kappa^{;\alpha}) \,-\,12\kappa
 \omega^{;\mu}\omega^{;\nu}\,-\,12\kappa^{;(\mu}\omega^{;\nu)}\,-\,2
 \kappa^{:\mu\nu}
\end{eqnarray}
and $\kappa = \omega_{;\alpha}\omega^{;\alpha}.$ 
The numerical coefficients as predicted by $\zeta-$function renormalization are given by
\begin{equation}
a\,=\,(2^{9} 45 \pi^{2})^{-1}\left[ 12 h(0)\,+\,18 h({1\over 2})\,+\,72 h(1)\right],
\end{equation}
\begin{equation}
b\,=\,(2^{9} 45 \pi^{2})^{-1}\left[ - 4 h(0)\,-\,11 h({1\over 2})\,-\,124 h(1)\right],
\end{equation}
and
\begin{equation}
c\,=\,- (2^{9} 45 \pi^{2})^{-1} 120 h(1),
\end{equation}
where $h(s)$ denotes the number of helicity states for fields of spin $s,$
while the dimensional renormalization gives
\begin{equation}
c(1)\,=\,0.
\end{equation}
Since the  transformational
rule for a general geometries is much more complicated  we restricted ourselves to the
Ricci-flat metrics.
The stress tensor in the optical space naturally splits into two parts:
\begin{equation}
<\tilde{T}^{\mu}_{\nu}>_{ren}\,=\,{\cal T}^{\mu}_{\nu}\,
+\,{9\,c\over 8\,M^{4}}x^{6}\delta^{\mu}_{0} \delta^{0}_{\nu},
\end{equation}
where $\cal{T}^{\mu}_{\nu}$ is a conserved traceless tensor and the second term in the
right hand side of (13) is constructed from the trace anomaly.

Now, we assume  $N=10$ and restrict ourselves to the scalar field. 
Hence, taking $\omega \,=1/2 \ln (|g_{tt}|)\,$ and making use of the regularity condition 
$|<T^{\theta}_{\theta}>| < \infty,$ one obtains
\begin{equation}
a_{9}\,=\,-\sum_{n =1}^{8}(10\,-\,n)a_{n},
\end{equation}
and
\begin{equation}
a_{10}\,=\,\sum_{n=1}^{8}(9 \,-\,n)a_{n}.
\end{equation}
Moreover, accepting the strong thermal hypothesis one concludes
that the coefficients $a_{i}$ for $1\,\leq\,i\,\leq\, 5$ should vanish.
Further, solving the conservation equation for $<\tilde{T}^{r}_{r}>$
in the optical space, 
transforming the resulting tensor back to the physical space and
employing the Christiensen-Fulling conditions, i. e. the conditions
which guarantee the regularity of the stress-energy tensor in the local frames
on the event horizon one has
\begin{equation}
< T_{\nu }^{\mu }>_{ren}\,=\,< T_{\nu }^{\mu }>^{Page}\,+\,\Delta^{\mu}_{\nu},
\end{equation}
where 
\begin{equation}
< T_{\nu }^{\mu }>^{Page}\,=\,T \left\{{1\,-\,x^{6} (4\,-\,3 x)^{2}\over (1\,-\,x)^{2}}
{\rm diag[-3,\,1,\,1,\,1]}^{\mu}_{\nu}\,+\,24 x^{6}
{\rm diag[3,\,1,\,0,\,0]}^{\mu}_{\nu}\right\},
\end{equation}
is the stress tensor evaluated within the framework
of the Page approximation~\cite{Page}, and the conserved and 
traceless tensor $\Delta^{\mu}_{\nu}$
is given by 
\begin{equation}
\Delta^{t}_{t}\,=
\,-3 T \left[{a_{6}\over 2} x^{6}\,+\,{1\over 366}(9 a_{6}\,-\,64 a_{8}) x^{7}\,-\,
{11\over 61}(3 a_{6}\,-\,a_{8}) x^{8}\right],
\end{equation}
\begin{equation}
\Delta^{r}_{r}\,=\,- T\left[ {a_{6}\over 2} x^{6}\,-\,{1\over 122}(13 a_{6}\,-\,16 a_{8}) x^{7}
\,-\,{9\over 61}(3 a_{6}\,-\,a_{8})x^{8}\right],
\end{equation}
and
\begin{equation}
\Delta^{\theta}_{\theta}\,=\,\Delta^{\phi}_{\phi}\,=\,
T \left[ a_{6} x^{6}\,-\,{1\over 61}( a_{6}\,+\,20 a_{8}) x^{7}\,-\,
{21\over 61}(3 a_{6}\,-\,a_{8}) x^{8}\right].
\end{equation}
Further determination of the model requires two pieces of numerical data.
Taking, for example, a horizon value of the tangential component of the stress
tensor (the radial component and hence $<T^{t}_{t}(1)>_{ren}$ may be easily obtained
from the trace anomaly) one gets
\begin{equation}
a_{6}\,=\,{1\over 3}(61\Theta \,-\,732\,-\,a_{8}),
\end{equation}
where $\Theta \,=\,1/T <T^{\theta}_{\theta}(1)>_{ren}.$
Finally, the remaining constant $a_{4}$ may be fixed by some sort of the best fit 
argument.
Here however we proceed differently: we perform the least-quare fit
to the available numerical data.
There are two published sources of information: the numerical estimates
carried out by Candelas and Howard~\cite{Candelas},\cite{Howard}, and more 
recently by Anderson, Hiscock, and Samuel~\cite{AHS1},\cite{AHS2}.
Although the latter authors presented their results only graphically
some of their results concerning the tangential 
component may by found in Ref.~\cite{Visser}.
In the region $[ 2 M,\,5 M]$ 
we adopted the Anderson,
Hiscock, and Samuel data as presented in Ref.~\cite{Visser},
whereas for $r\,>\,5 M$ we accept the results of  numerical 
calculations carried out by Howard~\cite{Howard}.
We discarded 3 points because the numerically determined 
trace exceeded the exact one by more than $1\%.$
Using Mathematica to perform the fit we obtained
\begin{equation}
a_{6}\,=\,-74.230,
\end{equation}
and
\begin{equation}
a_{7}\,=\,-329.135.
\end{equation}
The maximal deviation of the fitted curve does not exceed $0.7\%.$

It should be noted that the logarithmic term appearing in the solution of the 
conservation equation i.e.$<\tilde{T}^{r}_{r}>$  survives if the regularity condition
of $<T^{\theta}_{\theta}>_{ren}$ is imposed. However, since the coefficient
in front of the logarithmic term involves $a_{1}$ and $a_{2}$
such a term by the thermal hypotheses vanish.

We complete the  discussion of the scalar $N\,=\,10$ case 
by comparing obtained approximation
(16-20) to the model developed earlier.
Introducing a new set of parameters $\alpha_{4}$ and $A_{8}$ in place of 
$a_{6}$ and $a_{8}$ 
\begin{equation}
a_{6}\,=\,{17\over 12} A_{8}\,-\,2\beta,
\end{equation}
and
\begin{equation}
a_{8}\,=\,{469\over 96}A_{8}\,-\,6\beta,
\end{equation}
one obtains precisely the results of Ref.~\cite{J1}.
Indeed, inserting (24,25) into (17-20) after simple rearrangements one has 
\begin{equation}
8\pi^2\Delta^{t}_{t}\,=\,{\frac{M^2}{r^6}} \left( {\frac{\beta}{240}}\,+\,{%
\frac{17}{6}}\alpha_{4}\right)\,-\, {\frac{M^{3}}{r^{7}}} \left({\frac{433}{%
27}}\alpha_{4}\,+\,{\frac{\beta}{120}}\,+\,{\frac{4}{405}}A_{8}\right)\, +\,{%
\frac{M^{4}}{r^8}} \left({\frac{11}{540}} A_{8}\,+\,{\frac{385}{18}}%
\alpha_{4}\right),
\end{equation}
\begin{equation}
8\pi^2\Delta^{r}_{r}\,=\,{\frac{M^2}{r^6}} \left( {\frac{\beta }{720}}\,+\,{%
\frac{17 }{18}}\alpha_{4}\right)\,-\, {\frac{M^3}{r^7}} \left({\frac{121}{27}%
}\alpha_{4}\,+\,{\frac{\beta}{360}}\,+\,{\frac{A_{8}}{405}}\right) \,+\, {%
\frac{M^4}{r^8}} \left({\frac{35 }{6}}\alpha_{4}\,+\,{\frac{A_{8}}{180}}
\right),
\end{equation}
and
\begin{equation}
8\pi^2 \Delta^{\theta}_{\theta}\,=\,-{\frac{M^2 }{r^6}} \left({\frac{17}{9}}%
\alpha_{4}\,+\,{\frac{\beta}{360}}\right)\,+\,{\frac{M^3}{r^7}} \left({\frac{%
277}{27}}\alpha_{4}\,+\,{\frac{\beta }{180}}\,+\,{\frac{A_{8}}{162}}\right)
\,-\,{\frac{M^4}{r^8}} \left({\frac{245 }{18}}\alpha_{4}\,+\,{\frac{7}{540}}%
A_{8}\right),
\end{equation}
where $\alpha_{4}$ by (2) is given by
\begin{equation}
\alpha_{4}\,=\,-{1\over 960}A_{8}
\end{equation}
For completeness we write out the general expression for the field
fluctuation. 
\begin{equation}
< \phi^{2} >_{ren}\,=\, {T^{2}_{H}\over 12} (
1\,+\,x\,+\,x^2\,+\,x^3\,+\,\alpha_{4} x^4\,-\,\alpha_{4} x^5).
\end{equation}
which is necessary ingredient of the Hadamard regularization.

Now let us consider the consequences of the assumption that the curvature corrections
to the stress-energy tensor are of order $x^{3}.$
From the analyses carried out by Jensen and Ottewill~\cite{Bruce} we know that analytical
approximation of the  stress-energy tensor of the vector field 
satisfies the thermal condition in its weak form.
Let $\wp$ be the order of the polynomials that describe resulting stress-energy tensor.
Taking $N\,=\,10$  
and repeating the calculations for fields of arbitrary spin one obtains 
$\wp\,=\,8$ involving 5 unknown parameters.
On the other hand $N\,=\,8$ yields $\wp\,=\,6$ polynomials  with 3 undetermined constants.
To simplify our discussion, in the further analyses we take $N\,=\,8.$ 
Since in the optical space the trace anomaly  of the conformally coupled massless
vector field does not vanish one has to take into account an analog of the Zannias term.
After some algebra one finds for scalar, spinor, and vector fields
\begin{eqnarray}
<T^{t}_{t}>_{ren}\,&=&\,-3 p T\left\{
1 + 2\,x + 3\,{x^2} + 4\,{x^3} - {x^4}\,\left( {\frac{97}{3}} - {\frac{176\,\alpha}{9\,p}} -
     {\frac{64\,\beta}{3\,p}} + {\frac{32\,\gamma}{3\,p}} 
	  +\, {\frac{11\,a_{4}}{9}} + 
     {\frac{17\,a_{5}}{18}} + {\frac{a_{6}}{3}} \right)\right.\nonumber\\
&+&\,
	 {x^5}\,\left( -{\frac{730}{9}} + {\frac{1232\,\alpha}{27\,p}} + 
	      {\frac{448\,\beta}{9\,p}} - {\frac{224\,\gamma}{9\,p}} - {\frac{74\,a_{4}}{27}} - 
     {\frac{95\,a_{5}}{54}} - {\frac{7\,a_{6}}{9}} \right)\nonumber\\
&&\left.+\, {x^6}\,\left( -{\frac{1225}{9}} + {\frac{1448\,\alpha}{27\,p}} + 
     {\frac{520\,\beta}{9\,p}} - {\frac{224\,\gamma}{9\,p}} - {\frac{119\,a_{4}}{27}} - 
     {\frac{70\,a_{5}}{27}} - {\frac{7\,a_{6}}{9}} \right)  \right\},
\end{eqnarray}
\begin{eqnarray}
<T^{r}_{r}>_{ren}\,&=&\,
p T \left\{ 1 + 2\,x + 3\,{x^2} 
+ {x^3}\,\left( {\frac{460}{3}} - {\frac{704\,\alpha}{9\,p}} - 
     {\frac{256\,\beta}{3\,p}} + {\frac{128\,\gamma}{3\,p}} + {\frac{56\,a_{4}}{9}} + 
     {\frac{34\,a_{5}}{9}} + {\frac{4\,a_{6}}{3}} \right) \right.\nonumber\\ 
&+&\, {x^4}\,\left( {\frac{575}{3}} - {\frac{880\,\alpha}{9\,p}} - 
     {\frac{320\,\beta}{3\,p}} + {\frac{160\,\gamma}{3\,p}} + {\frac{61\,a_{4}}{9}} + 
     {\frac{85\,a_{5}}{18}} + {\frac{5\,a_{6}}{3}} \right)\nonumber\\ 
&+&\,{x^5}\,\left( {\frac{578}{3}} - {\frac{880\,\alpha}{9\,p}} - 
     {\frac{320\,\beta}{3\,p}} + {\frac{160\,\gamma}{3\,p}} + {\frac{58\,a_{4}}{9}} + 
     {\frac{73\,a_{5}}{18}} + {\frac{5\,a_{6}}{3}} \right) \nonumber\\ 
&&\left.+\, {x^6}\,\left( 175 - {\frac{248\,\alpha}{3\,p}} - 
     {\frac{88\,\beta}{p}} + {\frac{32\,\gamma}{p}} + {\frac{17\,a_{4}}{3}} + 
     {\frac{10\,a_{5}}{3}} + a_{6} \right)\right\}, 
	 \end{eqnarray}
and	 
\begin{eqnarray}
 <T^{\theta}_{\theta}>_{ren}\,&=&\,p T \left\{
 1 + 2\,x + 3\,{x^2} 
 + {x^3}\,\left( -{\frac{212}{3}} + {\frac{352\,\alpha}{9\,p}} + 
       {\frac{128\,\beta}{3\,p}} - {\frac{64\,\gamma}{3\,p}} - {\frac{28\,a_{4}}{9}} - 
     {\frac{17\,a_{5}}{9}} - {\frac{2\,a_{6}}{3}} \right)\right.\nonumber\\ 
&+& {x^4}\,\left( -{\frac{433}{3}} + {\frac{704\,\alpha}{9\,p}} + 
       {\frac{256\,\beta}{3\,p}} - {\frac{128\,\gamma}{3\,p}} - {\frac{47\,a_{4}}{9}} - 
       {\frac{34\,a_{5}}{9}} - {\frac{4\,a_{6}}{3}} \right)\nonumber\\  
&+& {x^5}\,\left( -218 + {\frac{352\,\alpha}{3\,p}} + 
      {\frac{128\,\beta}{p}} - {\frac{64\,\gamma}{p}} - {\frac{22\,a_{4}}{3}} - 
      {\frac{14\,a_{5}}{3}} - 2\,a_{6} \right) \nonumber\\ 
&&\left.+ {x^6}\,\left( -{\frac{875}{3}} + {\frac{1312\,\alpha}{9\,p}} + 
      {\frac{464\,\beta}{3\,p}} - {\frac{160\,\gamma}{3\,p}} - {\frac{85\,a_{4}}{9}} - 
      {\frac{50\,a_{5}}{9}} - {\frac{5\,a_{6}}{3}} \right) \right\},
\end{eqnarray}
where $\alpha\,=\,a (64 M^{4} T)^{-1},$ $\beta\,=\,b  (64 M^{4} T)^{-1},$ and
$\gamma\,=\, c  (64 M^{4} T)^{-1}.$
Surprisingly $< T^{t}_{t}>_{ren}$ 
is of the form (4) with $m\,=\,3.$
It could be shown that similar approximate stress-energy tensor 
may be obtained from the formulae
derived recently by Visser~\cite{Visser}.

Inspection of (31-33) shows that
for  $N\leq\,8$ the result is described by polynomials of order 6, because of
the geometrical terms that contribute to  $<T^{\mu}_{\nu}>_{ren}.$ 
Indeed, for example making use of the additional 
constraints (obtained from equations $a_{7}\,=\,0$
and $a_{8}\,=\,0$)
\begin{equation}
a_{5}\,=\,- 66 \,+\,{\frac{35\,\alpha}{p}} + {\frac{39\,\beta}{p}} - 
{\frac{39\,\gamma}{2\,p}} - 2\,a_{4}
\end{equation}
\begin{equation}
a_{6}\,=\,45 - {\frac{47\,\alpha}{2\,p}} - {\frac{51\,\beta}{2\,p}} + 
{\frac{51\,\gamma}{4\,p}} + a_{4}
\end{equation}
one obtains a simple stress-energy tensor that depends on undetermined 
parameter $a_{4}.$ \cite{J2}-\cite{J4}
The free parameter may be fixed from the known value of one component of the 
stress- energy tensor, say,  $< T^{\theta}_{\theta}>_{ren}$ on the event horizon. 
Taking, for example, in the vector case, 
\begin{equation}
 < T^{\theta}_{\theta}>_{ren}\,=\,-{1\over 240 \pi^{2} M^{4}}
\end{equation}
results in the approximation that coincides with the analytic 
part of the Jensen and Ottewill evaluation of the stress-energy tensor~\cite{Bruce},\cite{J4}.
The interesting property of
$N\,=\,6$ model is that the difference $< T^{t}_{t}>_{ren}\,-\,< T^{r}_{r}>_{ren}$ 
does not depend
on the parameter $a_{4}$ and consequently the entropy of quantized fields could be 
constructed~~\cite{J2}.

To this end, we report the results of our $N = 11$ calculations 
which generalize Eqs. (16 - 20). 
The stress tensor of the scalar field is  described by ${\wp}\,=\,9$ 
polynomials with 3 free parameters and has a general form (16) with 
\begin{eqnarray}
{1\over T}\Delta^{\theta}_{\theta}\,&=&\,{1\over 17}\left( 5844\,-\,487 \Theta\,-17 a_{9}\,-
\,27 a_{10}\right) x^{6}\,+\,{1\over 17}\left( 252 \Theta\,-\,3024\,+\,17 a_{9}\,+\,22 a_{10}
\right) x^{7}\nonumber\\
&+&\,{3\over 17}\left[ 56 \Theta\,+\,3\left( a_{10}\,-\,224\right)\right] x^{8}\,+\,
{4\over 17 } \left( 21 \Theta \,-\,252\,-\,a_{10}\right) x^{9},
\end{eqnarray}

\begin{eqnarray}
{1\over T}\Delta^{r}_{r}\,&=&\,{1\over 34}
\left( 487 \Theta\,-\,5844\,+\,17 a_{9}\,+\,27 a_{10}\right) x^{6}\,+\,
{1\over 34}\left( 3588\,-\,299 \Theta\,-\,17 a_{9}\,-\,23 a_{10}\right) x^{7}
\nonumber\\
&+&\,{1\over 34}\left( 1908\,-\,159 \Theta\,-\,7 a_{10}\right) x^{8}\,+\,
{3\over 34}\left( 252\,-\,21 \Theta\,+\,a_{10}\right) x^{9},
\end{eqnarray}
and
\begin{eqnarray}
-{1\over 3 T}\Delta^{t}_{t}\,&=&\,{1\over 34}
\left(5844\,-\,487 \Theta\,-\,17 a_{9}\,-\,27 a_{10}\right) x^{6}\,+\,
{1\over 102}\left( 709 \Theta\,-\,8508\,+\,51 a_{9}\,+\,65 a_{10}\right) x^{7}
\nonumber\\
&+&\,{1\over 102}\left( 513 \Theta\,-\,6156\,+\,29 a_{10}\right) x^{8}\,+\,
{13\over 102}\left( 21\Theta\,-\,252\,-\,a_{10}\right) x^{9}.
\end{eqnarray}
Although generalization of the method to $N > 11$ is obvious 
it seems that such models are of little use.

Final determination of the model is achieved by performing 
the least square fit of the $\langle T^{\theta}_{\theta}\rangle_{ren}$
to the  numerical data discussed earlier. 
Using Mathematica once more to perform the linear least-square fit we find
\begin{equation}
a_{9}\,=\,123.347
\end{equation}
\begin{equation}
a_{10}\,=\,-1.071
\end{equation}
and
\begin{equation}
\Theta\,=\,10.245.
\end{equation}
We have also fitted the curve to the Anderson et al. data alone
and obtained similar results.
Note that one of the parameters is the horizon value of the tangential
component of the stress-energy tensor.
The maximal deviation of the fitted curve does not exceed $0.7\%.$
On the other hand however, one should remember that the accuracy 
of the numerical data is limited and as in other calculations of this 
type we should not trust (16) with (37-39) beyond $1\%$ level.



\end{document}